\documentclass[twocolumn,superscriptaddress,aps,10pt]{revtex4}
\usepackage{amsfonts}
\usepackage{amsmath}
\usepackage{amssymb}
\usepackage{graphicx}
\usepackage{dcolumn}
\usepackage{times}

\DeclareMathOperator{\sech}{sech}

\newcommand{\intl}{\int_{-\infty}^{+\infty}}

\begin{document}

\title{Statics and dynamics of atomic dark-bright solitons in the presence of delta-like impurities}
\author{V.~Achilleos}
\affiliation{Department of Physics, University of Athens, Panepistimiopolis,
Zografos, Athens 157 84, Greece}
\author{P.G.~Kevrekidis}
\affiliation{Department of Mathematics and Statistics, University of Massachusetts,
Amherst, Massachusetts 01003-4515, USA}
\author{V.M.~Rothos}
\affiliation{Department of Mathematics, Physics Computational
Sciences, Faculty of Engineering, Aristotle University of Thessaloniki, Thessaloniki 54124, Greece}
\author{D.J.~Frantzeskakis}
\affiliation{Department of Physics, University of Athens, Panepistimiopolis,
Zografos, Athens 157 84, Greece}

\begin{abstract}

Adopting a mean-field description for a two-component atomic Bose-Einstein condensate, we study the statics and dynamics of dark-bright solitons in the presence of localized impurities. We use adiabatic perturbation theory to derive an equation of motion for the dark-bright soliton center. We show that, counter-intuitively, an attractive (repulsive) delta-like impurity, acting solely on the bright soliton component, induces an effective localized barrier (well) in the effective potential felt by the soliton; this way, dark-bright solitons are reflected from (transmitted through) attractive (repulsive) impurities. Our analytical results for the small-amplitude oscillations of solitons are found to be in good agreement with results obtained via a Bogoliubov-de Gennes analysis and direct numerical simulations.

\end{abstract}

\pacs{05.45.Yv,~03.75.Mn,~03.75.Kk}

\maketitle

\section{Introduction}

The physics of atomic Bose-Einstein condensates (BECs) \cite{book1,book2} has offered the possibility of the study of purely nonlinear phenomena in the mesoscopic scale. In particular, there has been a vast amount of research efforts devoted to the study of macroscopic nonlinear excitations of BECs (see, e.g., Refs.~\cite{emergent,revnonlin,fetter,rab,djf} for reviews on this topic). In that regard, of particular interest are the so-called matter-wave solitons, of either the bright \cite{rab} or the dark \cite{djf} type, that can be supported in BECs with attractive or repulsive interactions, respectively. Nevertheless, these types of solitons may coexist in multi-component condensates with repulsive interactions (see, e.g., Refs.~\cite{BA} and \cite{DDB} for relevant work in two-component and spinor condensates, respectively): such, so-called dark-bright (DB) solitons exist due to the fact that the dark-soliton component creates, through the inter-species interaction, a trapping mechanism that allows the bright soliton component to be formed (even though this is not possible in repulsive BECs). Dark-bright solitons have been studied extensively in other contexts, such as nonlinear optics \cite{yuri} and the theory of nonlinear waves \cite{ablowitz}. Furthermore, they have recently been analyzed in discrete settings \cite{azucena}, while higher-dimensional generalizations ---namely, vortex-bright-soliton structures--- were studied as well~\cite{VB}. Importantly, dark-bright solitons have been observed in experiments conducted both in optics \cite{seg1,seg2} and, more recently, in BECs \cite{hamburg,engels1,engels2,engels3}.

On the other hand, the interaction of solitons with localized impurities is a quite general and fundamental problem that
has attracted much attention in the theory of nonlinear waves \cite{RMP} and solid state physics \cite{marad,kos1}. In this context, the interaction of either bright or dark solitons with $\delta$-like impurities has been investigated in the framework of the nonlinear Schr\"{o}dinger (NLS) equation (see, e.g., Refs.~\cite{kos2,cao,holmes,vvk1}, while relevant studies have also appeared in the physics of atomic BECs (see, e.g., \cite{gt,nn,greg,leebrand,brand2}). In the latter context, localized impurities may easily be created as sharply-focused far-detuned laser beams and can be used to manipulate matter-wave dynamics (see, e.g., Ch.~17 of Ref.~\cite{emergent}), while they have already been used in experiments for the creation of solitons \cite{engels,hulet}. Nevertheless, to the best of our knowledge, the problem of the interaction of matter-wave dark-bright solitons with localized impurities has not been addressed so far.

In this work, we aim to study this problem in the framework of mean-field theory. More specifically, we consider a quasi one-dimensional (1D) two-component repulsive BEC, composed by two hyperfine states of the same alcali species (as, e.g., in the experiments of Refs.~\cite{engels1,engels2,engels3}) --- a system that can be approximated by two coupled 1D Gross-Pitaevskii  equations (GPEs) (see, e.g., Refs.~\cite{emergent,revnonlin,djf}). We assume that both components are confined by the usual harmonic trap, while an additional small-amplitude localized ($\delta$-like) impurity potential is also incorporated in both components. We employ the hamiltonian approach of the perturbation theory of matter-wave solitons (see, e.g., Ref.~\cite{djf}) to study analytically the adiabatic dynamics of DB solitons supported in the system. This way, we derive an effective equation of motion for the DB-soliton center. We find that if the impurity potential acts solely on the bright-soliton component then
the soliton-impurity interaction is effectively repulsive (attractive) for a genuinely attractive (repulsive) impurity. This behavior is in a sharp contrast with the one corresponding to dark solitons in single-component condensates: there, the nature of the dark soliton---impurity interaction is the same as
the type of the impurity (i.e., repulsive/attractive for repulsive/attractive
impurities, respectively) \cite{gt}.

We study the statics and dynamics of solitons near the fixed points of the effective potential associated to the above mentioned equation of motion using both our analytical approach and numerical simulations. We also perform a Bogoliubov - de Gennes (BdG) analysis to investigate the excitation spectra of the DB-solitons proper and study their stability. Where appropriate, we find a very good agreement between the analytical predictions and the numerical findings, e.g., the characteristic frequencies obtained by the equation of motion and the eigenfrequencies of the internal modes (also known as ``anomalous modes'' \cite{book2,fetter}) associated with DB-solitons.

The paper is structured as follows. In Section 2 we present the model, use perturbation theory, and derive the equation of motion for the soliton center. In section 3, we analyze the effective potential and forces acting on the dark-bright solitons and identify the most interesting case, i.e., when the impurity acts solely in the bright-soliton component. Section 4 is devoted to a systematic comparison of our analytical findings with simulations, including results of the BdG analysis. Finally, in section 5 we summarize our conclusions.

\section{Model and analytical considerations}

\subsection{Setup}
We consider a two-component elongated (along the $x$-direction) repulsive BEC, composed of two different
hyperfine states of the same alkali isotope. Assuming that the trap is highly anisotropic, with the longitudinal and transverse trapping frequencies being such that $\omega_x \ll \omega_{\perp}$, we may describe this system by the following two coupled GPEs \cite{emergent,revnonlin}:
\begin{eqnarray}
i\hbar \partial_t \psi_j =
\left( -\frac{\hbar^2}{2m} \partial_{x}^2 +V_j(x) -\mu_j + \sum_{k=1}^2 g_{jk} |\psi_k|^2\right)\!\psi_j.
\label{model}
\end{eqnarray}
Here, $\psi_j(x,t)$ ($j=1,2$) denote the mean-field wave functions of the two components (normalized to the
numbers of atoms $N_j = \int_{-\infty}^{+\infty} |\psi_j|^2 dx$), $m$ is the atomic mass, $\mu_j$ are the chemical potentials, $g_{jk}=2\hbar\omega_{\perp} a_{jk}$ are the effective 1D coupling constants,
$a_{jk}$ denote the three $s$-wave scattering lengths (note that $a_{12}=a_{21}$) that account for collisions between atoms belonging to the same ($a_{jj}$) or different ($a_{jk}, j \ne k$) species, and $V_j(x)$ represent the external trapping
potentials.

We assume that both components are confined by the usual harmonic trap, namely
$V(x)=(1/2)m\omega_x^2 x^2$, while an additional localized ``impurity'' potential, which may be created by a far-detuned laser beam, is also present. If such an impurity is strongly localized, one may theoretically approximate its spatial profile by a $\delta$-function; thus, the trapping potentials for the two components can be described as: $V_j(x)=V(x) + b_j \delta(x)$, where $b_j$ are the barrier amplitudes in each component. Note that for a blue- or red-detuned laser beam, the impurity potential can either repel ($b_j>0$) or attract ($b_j<0$) the atoms of the respective component of the condensate.

We examine the case where the two-component BEC under consideration consists of two different hyperfine states
of $^{87}$Rb, such as the states $|1,-1\rangle$ and $|2,1\rangle$ used in the experiment of Ref.~\cite{mertes},
or the states $|1,-1\rangle$ and $|2,-2\rangle$ used in the experiments of Refs.~\cite{engels1,engels2,engels3}. In the first case the scattering lengths take the values $a_{11}=100.4a_0$, $a_{12}=97.66a_0$ and $a_{22}=95.00a_0$, while in
the second case the respective values are $a_{11}=100.4a_0$, $a_{12}=98.98a_0$ and $a_{22}=98.98a_0$ (where
$a_0$ is the Bohr radius). In either case, the scattering lengths take approximately the same
values, say $a_{ij} \approx a$, which is what we will assume hereafter. Thus, measuring the densities $|\psi_j|^2$, length, time and energy in
units of $2a$, $a_{\perp} = \sqrt{\hbar/\omega_{\perp}}$, $\omega_{\perp}^{-1}$ and $\hbar\omega_{\perp}$,
respectively, we may cast Eqs.~(\ref{model}) into the following dimensionless form,
\begin{eqnarray}
i \partial_t u_d  =& -&\frac{1}{2} \partial_{x}^2u_d  + V_d(x)u_d \nonumber \\
&+&(|u_d|^2 + |u_b|^2 -\mu) u_d,
\label{deq1}
\\
i \partial_t u_b  =& -&\frac{1}{2} \partial_{x}^2u_b +V_b(x)u_b \nonumber \\
&+& (|u_b|^2 + |u_d|^2- \mu-\Delta) u_b.
\label{deq2}
\end{eqnarray}
In the above equations, we have used the notation $\psi_1 = u_d$ and $\psi_2 = u_b$, indicating that the component $1$ ($2$) will be supporting a dark (bright) soliton. Notice that the respective normalized chemical potentials read
$\mu_1=\mu_d=\mu$ and $\mu_2=\mu_b=\mu+\Delta$, and below we will assume that $\mu_d>\mu_b$ (i.e., $\Delta =-|\Delta|<0$). Finally, the external potentials in Eqs.~(\ref{deq1})-(\ref{deq2}) take the form
\begin{eqnarray}
V_d(x)&=&V(x)+b_1\delta(x) = \frac{1}{2}\Omega^{2} x^{2}+b_1\delta(x)  \label{vd} \\
V_b(x)&=&V(x)+b_2\delta(x) = \frac{1}{2}\Omega^{2} x^{2}+b_2\delta(x),
\label{vb}
\end{eqnarray}
where $\Omega = \omega_x/\omega_\perp $ and $b_1$, $b_2$ are the normalized trap strength and barrier prefactor strength, respectively. Below, both of
these parameters will be considered to be small, i.e., $\Omega \sim b \ll 1$.

Before proceeding further, it is necessary to consider at first the effect of the impurity on the Thomas-Fermi (TF) cloud carrying the dark soliton. According to the analysis of Ref.~\cite{gt}, the TF density near the trap center (where the impurity is located) can be approximated as:
\begin{eqnarray}
|u_{\rm TF}|^2 & \approx &\mu-2\sqrt{\mu} f(x),
\label{tf1} \\
f(x)&=&\frac{\Omega^2}{4\sqrt{\mu}}x^2 + \frac{b_1}{2}\exp(-2\sqrt{\mu}|x|),
\label{tf2}
\end{eqnarray}
where $f(x)$ is considered to be small with respect to the chemical potential $\mu$. The first term in the right-hand side of Eq.~(\ref{tf2}) accounts for the unperturbed TF density (in the absence of the impurity); on the other hand, the second term
actually approximates the delta-like impurity, which creates in the TF density a localized dip (hump) for $b_1<0$ ($b_1>0$); the latter has obviously a discontinuous derivative at $x=0$ due to the matching conditions at $x=0$ (see details in
Ref.~\cite{gt}).


\subsection{Perturbation theory}

We assume that the dark soliton is on top of a modified TF cloud, as described by Eqs.~(\ref{tf1})-(\ref{tf2}). Accordingly, the density $|u_d|^2$ in Eqs.~(\ref{deq1})-(\ref{deq2}) is substituted by $|u_d|^2 \rightarrow |u_{\rm TF}|^2 |u_d|^2$. Furthermore, introducing the transformations $t \rightarrow \mu t$, $x \rightarrow {\sqrt{\mu}}x$, $|u_b|^2 \rightarrow \mu^{-1} |u_b|^2$, we cast Eqs.~(\ref{deq1})-(\ref{deq2}) into the form:
\begin{eqnarray}
&&i \partial_{t}u_d  +\frac{1}{2} \partial_{x}^2 u_d  -(|u_d|^2 +  |u_b|^2 -1) u_d = R_d,
\label{eq1d} \\
&&i \partial_{t} u_b  +\frac{1}{2} \partial_{x}^2u_b - (|u_d|^2 + |u_b|^2-\tilde{\mu}) u_b = R_b,
\label{eq2d}
\end{eqnarray}
where $\tilde{\mu} = 1+\Delta/\mu$, and
\begin{eqnarray}
R_d &\equiv& (2\mu^2)^{-1}\left[2(1-|u_d|^2)V(x)u_d +V'(x)\partial_x u_d \right] \nonumber \\
&+& b_1\mu^{-1/2}\left[\left(1-|u_d|^2\right)u_d-\frac{x}{|x|}\partial_x u_d\right]{\rm e}^{-2|x|},
\label{Rd} \\
R_b &\equiv &\mu^{-2}\Big[(1-|u_d|^2)V(x)u_b+ b_2 \mu \delta(x)u_b \nonumber \\
 &-& b_1\mu^{3/2} |u_d|^2 u_b {\rm e}^{-2|x|}  \Big],
\label{Rb}
\end{eqnarray}
with $V'(x)\equiv dV/dx$. Equations (\ref{eq1d})-(\ref{eq2d}) can be viewed as a system of two coupled perturbed NLS equations, with perturbations given by Eqs.~(\ref{Rd})-(\ref{Rb}). In the absence of the perturbations ($\Omega=0$, $b_{1,2}=0$), and considering the boundary conditions $|u_d|^2 \rightarrow 1 $ and $|u_b|^2 \rightarrow 0$ as $|x| \rightarrow \infty$, the NLS Eqs.~(\ref{eq1d})-(\ref{eq2d}) possess an exact analytical DB soliton solution of the following form (see, e.g., Ref.~\cite{BA}):
\begin{eqnarray}
\!\!\!\!\!\!
u_d(x,t)&=&\cos\phi\tanh\left[D(x-x_0(t))\right]+i\sin\phi,
\label{dsoliton2}
\\
\!\!\!\!\!\!
u_b(x,t)&=&\eta\sech\left[D(x-x_0(t))\right]\exp\left[ikx+i\theta(t)\right],
\label{bsoliton2}
\end{eqnarray}
where $\phi$ is the dark soliton's phase angle, $\cos\phi$ and $\eta$ represent the amplitudes of the dark and
bright solitons, $D$ and $x_0(t)$ denote the width and the center of the DB soliton, while
$k=D\tan\phi = {\rm const}$ and $\theta(t)$ are the wavenumber and phase of the bright soliton,
respectively. The above parameters of the DB-soliton are connected through the following equations:
\begin{eqnarray}
D^2&=& \cos^2\phi-\eta^2,
\label{width} \\
\dot{x}_0 &=& D\tan\phi,
\label{x0} \\
\theta(t)&=&\frac{1}{2}(D^2-k^2)t+(\Delta/\mu)t,
\label{omegat}
\end{eqnarray}
where $\dot{x}_0$ is the DB soliton velocity. Notice that the amplitude $\eta$ of the bright soliton, the dark-soliton component's chemical potential $\mu$, as well as the width $D$ of the DB-soliton are connected
with the number of atoms of the bright soliton by means of the following equation
[for the variables appearing in Eqs.~(\ref{deq1})-(\ref{deq2})]:
\begin{equation}
N_b\equiv\intl |u_b|^2dx=\frac{2\sqrt{\mu}\eta^2}{D}.
\label{Nbmod}
\end{equation}
Let us now assume that the DB-soliton evolves adiabatically in the presence of the small perturbation, and employ the Hamiltonian approach of the perturbation theory for matter-wave solitons (see, e.g., Refs.~\cite{revnonlin,djf}) to study the DB-soliton dynamics. We start by considering the Hamiltonian (total energy) of the system of Eqs.~(\ref{eq1d})-(\ref{eq2d}), when the perturbations are absent ($R_d=R_b=0$), namely,
\begin{eqnarray}
E &=& \frac{1}{2}\int_{-\infty}^{+\infty} \mathcal{E} dx, \nonumber \\
\mathcal{E} &=& |\partial_{x} u_d|^2+|\partial_{x} u_b|^2+(|u_d|^2+|u_b|^2-1)^2 \nonumber \\
&-&2(\Delta/\mu)|u_b|^2.
\label{energy}
\end{eqnarray}
The energy of the system, when calculated for the DB-soliton solution of Eqs.~(\ref{dsoliton2})-(\ref{bsoliton2}), takes the following form:
\begin{eqnarray}
E=\frac{4}{3}D^3+\chi \left(\frac{1}{2}D^2\sec^2\phi-\frac{\Delta}{\mu}\right), \quad
\chi \equiv \frac{N_b}{\sqrt{\mu}}.
\label{chi}
\end{eqnarray}
Since we have considered an adiabatic evolution of the DB soliton, we may assume that, in the presence of the perturbations of Eqs.~(\ref{Rd})-(\ref{Rb}), the DB soliton parameters become slowly-varying unknown functions of time $t$ (see, e.g., \cite{djf}). Thus, the DB soliton parameters become $\phi \rightarrow \phi(t)$, $D \rightarrow D(t)$, and, as a result,
Eqs.~(\ref{width})-(\ref{x0}) read:
\begin{eqnarray}
D^2(t)&=&\cos^2\phi(t)-\frac{1}{2} \chi D(t),
\label{s1} \\
\dot{x}_0(t)&=&D(t)\tan\phi(t),
\label{s2}
\end{eqnarray}
where we have used Eq.~(\ref{Nbmod}). The evolution of the parameters $\phi(t)$, $D(t)$
and $x_0(t)$ can be found by means of the evolution of the DB soliton energy.
In particular, employing Eq.~(\ref{chi}), it is readily found that
\begin{equation}
\frac{dE}{dt}=4\dot{D}D^2+\chi D\sec^2\phi(\dot{D}+D\dot{\phi} \tan\phi).
\label{denergy1}
\end{equation}
On the other hand, using Eqs.~(\ref{eq1d})-(\ref{eq2d}) and their complex conjugates, it can be found that
the evolution of the DB soliton energy, due to the presence of the perturbations, is given by:
\begin{equation}
\frac{dE}{dt}=-2{\rm Re} \left\{ \int_{-\infty}^{+\infty} \left( R_d^{\ast}\partial_t u_d + R_b^{\ast}\partial_t u_b \right) dx \right\},
\label{perturb}
\end{equation}
where asterisk denotes complex conjugate. Substituting $R_d$ and $R_b$ [cf.~Eqs.~(\ref{Rd})-(\ref{Rb})] into Eq.~(\ref{perturb}) and evaluating the integrals, we obtain from Eqs.~(\ref{s1}), (\ref{s2}), (\ref{denergy1}) and Eq.~(\ref{perturb}) a system of three equations for the evolution of the soliton parameters $\phi(t)$, $D(t)$
and $x_0(t)$. This system is linearized around its fixed point (see details in the Appendix) and,
in the physically relevant case of sufficiently small $\chi$, leads to the following  equation of motion for the small-amplitude displacement $X_0$ of the soliton position from the trap center:
\begin{eqnarray}
\ddot{X}_0 &=&-\frac{\partial V_{\rm eff}}{\partial X_0},
\label{eqmot}
\end{eqnarray}
where we have used the variables used in Eqs.~(\ref{deq1})-(\ref{deq2})). The effective potential in Eq.~(\ref{eqmot}) is given by
\begin{eqnarray}
V_{\rm eff}(X_0)&=&\frac{1}{2}\omega_{\rm osc}^2 X_0^2+ b\sech^2(D_0X_0),
\label{pot}
\end{eqnarray}
where the oscillation frequency $\omega_{\rm osc}$ and the parameter $b$ are respectively given by:
\begin{eqnarray}
\omega^2_{\rm osc}&=&\Omega^2 \left(\frac{1}{2}-\frac{\chi}{8\sqrt{1+\left(\frac{\chi}{4}\right)^2}}\right),
\label{omegaeff}\\
b&=&\frac{1}{6\left[8D_0\tilde{D}_0+\chi(2\tilde{D}_0-D_0)\right]}
\nonumber \\
&\times& \left[2 \left(1+2D_0^2\right)b_1  + \chi D_0 b_1 - 3 \chi D_0^2 b_2 \right],
\label{va2}
\end{eqnarray}
and $D_0$ and $\tilde{D}_0$ are constants of order $O(1)$ (see Appendix). Equation~(\ref{eqmot}) has the form of an equation of motion for a classical particle, with the coordinate $X_0$, moving in the effective potential $V_{\rm eff}$. Note that in the absence of the impurities [$b_1=b_2=0$, i.e., $b=0$ in Eq.~(\ref{pot})], Eq.~(\ref{eqmot}) recovers the results of Ref.~\cite{BA}: according to this work, a DB-soliton oscillates in a harmonic trap of strength $\Omega$ with the frequency $\omega_{\rm osc}$, given in Eq.~(\ref{omegaeff}); this frequency depends on the parameter $\chi$, i.e., the number of atoms $N_b$ of the bright soliton [see the definition of $\chi$ in Eq.~(\ref{chi})]. Below, we analyze the more general case, studying the effect of the impurities on the statics and dynamics of DB-solitons.

\section{The effective potential and forces}

The part of the effective potential (\ref{pot}) induced by the impurities consists of three different terms, as seen by the expression of the constant $b$ in Eq.~(\ref{va2}). Taking into regard that $D_0$ and $\tilde{D}_0$ are of order $O(1)$ and the parameter $\chi$ is small (as mentioned above -- see also the Appendix), it is readily observed that the sign of the
parameter $b$ is mainly determined by the leading-order term, $\propto 2 b_1 \left(1+2D_0^2\right)$, in Eq.~(\ref{va2}).
Thus, it is clear that the term $\propto \sech^2(D_0X_0)$ in the effective potential (\ref{pot}) is either a localized barrier (for $b_1>0$) or a localized well (for $b_1<0$). Here we should note that although Eq.~(\ref{pot}) is formally valid for small $\chi$, a numerical investigation of the more general case corresponding to values of $\chi$ of order $O(1)$ reveals that the nature of the potential is correctly captured by the above analysis. This can be understood by the fact that, generally speaking, increase of $\chi$ results in a decrease of $D_0$ from its maximum value (which is $D_0=1$) [see Eq.~(\ref{fixedpoint}) in the Appendix] and, thus, the sign of $b$ is always determined by the sign of $b_1$.

\begin{figure}[tbp]
\includegraphics[scale=0.25]{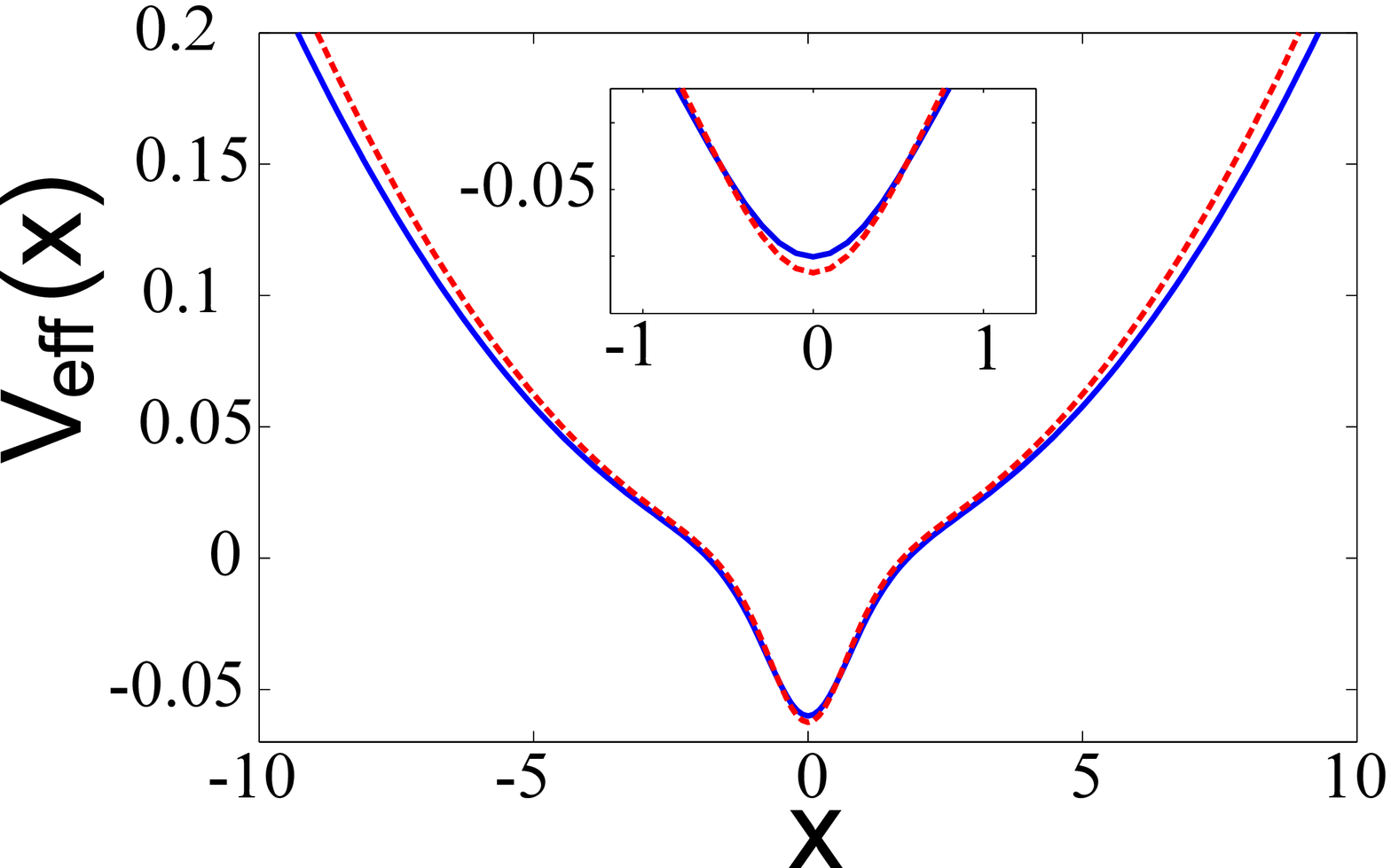}
\includegraphics[scale=0.25]{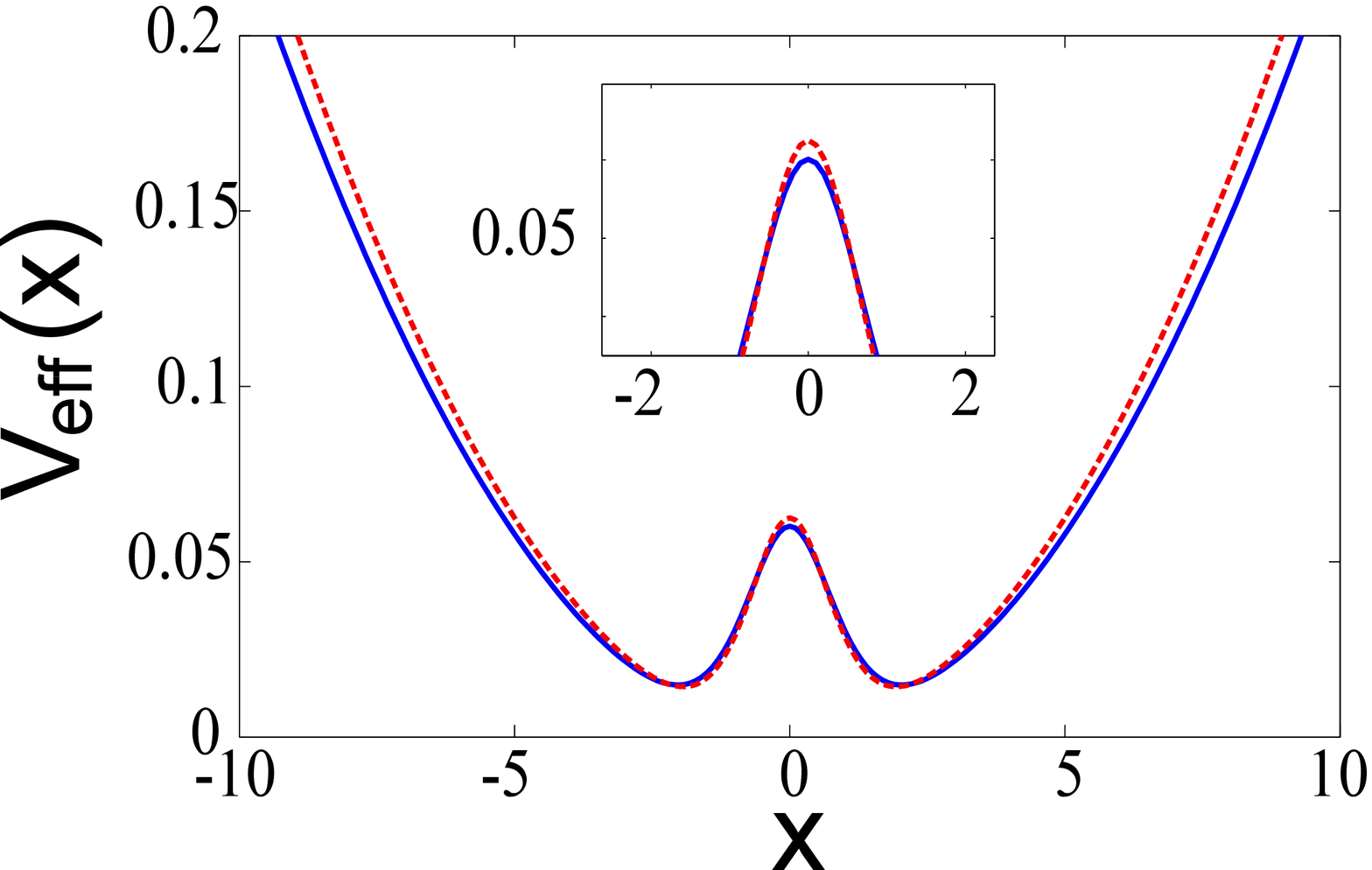}
\caption{(Color online) The effective potential (\ref{pot}) in the cases $\chi=0.7$ [solid (blue) line] and $\chi=0$ [dashed (red) line], i.e., in the absence of the bright-soliton component. The top and bottom panels correspond to $b_1 = b_2 = -0.15$ and $b_1 = b_2 = 0.15$, respectively; the harmonic trap strength is $\Omega=0.1$. Insets show details of the effective potentials in these cases near the trap center (where the impurities are located).}
\label{vefft}
\end{figure}

This result suggests that the form of the effective potential is not significantly modified due to the presence of the bright soliton component, as shown in Fig.~\ref{vefft}. Furthermore, numerical simulations in the framework of the GP Eqs.~(\ref{deq1})-(\ref{deq2}) (not shown here) for the DB soliton dynamics confirm the above picture. We have found that stationary DB-soliton states exist at the fixed points of the effective potential, and if these stationary states are displaced, they perform oscillations in the effective potential shown in the top panel Fig.~\ref{vefft} or, depending on their initial energy, they are either reflected or transmitted from the effective barrier in the bottom panel of Fig.~\ref{vefft}. This behavior was already described in detail in Ref.~\cite{gt}, where the interaction of matter-wave dark solitons with localized impurities was studied, hence we will not discuss it further.

\begin{figure}[tbp]
\centering
\includegraphics[scale=0.35]{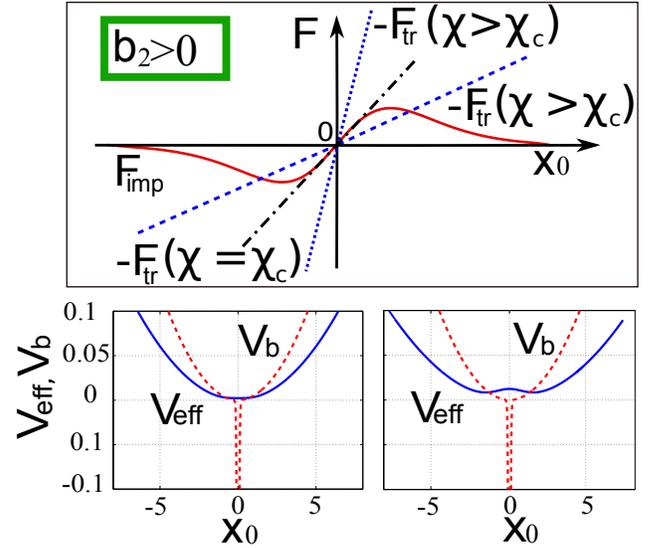}
\caption{(Color online) Top panel: A representative illustration of the dependence of the two forces acting on the soliton, $F_{\rm tr}$ and $F_{\rm imp}$ [solid (red) line], on the soliton center $X_0$; the impurity is assumed to be attractive, i.e., $b_2<0$. The dotted (blue) and dashed-dotted (black) lines show $-F_{\rm tr}$ for $\chi \le \chi_c$; in this case, the only fixed point is $X_0^{\ast}=0$. The dashed (blue) line shows $-F_{\rm tr}$ for $\chi >\chi_c$; in this case, there exist three fixed points. Note that since the profile $F_{\rm imp}$ remains qualitatively the same as  $\chi$ changes, for simplicity of illustration, it is plotted only for a single value of $\chi$. Bottom panels: the effective potential of Eq.~(\ref{pot}) [solid (blue) line], is plotted as a function of $X_0$, for $\chi=0.13<\chi_c=0.145$ (left) and $\chi=1.3> \chi_c=0.145$ (right), and it is compared to the actual potential $V_b$ acting on the bright component [cf. Eq.~(\ref{vb})], indicated by the dashed (red) line. Parameter values are $\mu=1$ and $\Omega=0.1$.}
\label{veff}
\end{figure}
\begin{figure}[tbp]
\centering
\includegraphics[scale=0.35]{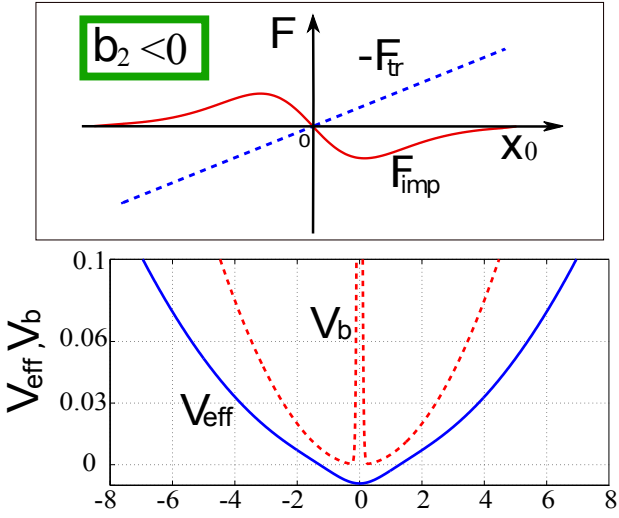}
\caption{(Color online) Similar to Fig.~\ref{veff}, but in the case of a repulsive barrier, $b_2>0$. In this case, the only fixed point is $X_0^{\star}=0$, as shown by the profiles of the forces (top panel). On the other hand, the effective potential (bottom panel) is always attractive.}
\label{veff1}
\end{figure}

Below, we focus on a quite interesting situation occurring if the impurity acts {\it solely} on the bright soliton component, i.e., $b_1=0$ and $b_2 \ne 0$: in this case, according to Eq.~(\ref{pot}), the forces acting on the
DB-soliton, i.e., the force exerted by the harmonic trap, $F_{\rm tr}$, and the localized impurity, $F_{\rm imp}$, read:
%
\begin{eqnarray}
F_{\rm tr}&=&-\omega_{\rm osc}^2 X_0,
\label{ftr}\\
F_{\rm imp}&=&2aD_0\sech^2(D_0X_0)\tanh(D_0X_0),
\label{fdef}
\end{eqnarray}
and the constant $a$, which is equal to $b$ for $b_1=0$, is given by:
\begin{eqnarray}
a&=&-\frac{ \chi D_0^2 b_2 }
{2\left[8D_0\tilde{D}_0+\chi(2\tilde{D}_0-D_0)\right]}.
\label{defina}
\end{eqnarray}
In this case, it can readily be observed that the parameter $a$ is negative (positive) when $b>0$ ($b<0$), as the denominator is positive for any value of $\chi$. This result is somehow counter-intuitive compared to the case where both impurities are present: now, if the impurity is repulsive then the respective force is attractive ($F_{\rm imp}<0$), while when it is attractive, the force is repulsive ($F_{\rm imp}>0$). Physically speaking, this behaviour can be understood by the nature of atom-impurity and atom-atom interactions: for example, if $b_2<0$ (attractive impurity) then the impurity attracts atoms in the bright soliton component; nevertheless, since inter-atomic interactions are repulsive, attracted atoms in $u_b$ repel atoms in the dark soliton component $u_d$. If the number of atoms in the bright component is sufficiently large, then the effect of the attractive impurity on the system is to create a repulsive potential
which overcomes the attractive effect of the trap yielding a repulsive
effective potential,
as shown in the bottom right panel of Fig.~\ref{veff}.


The form of the effective forces, $F_{\rm tr}$ and $F_{\rm imp}>0$, is illustrated in Figs.~\ref{veff} and \ref{veff1}:
there, the dependence of the forces on the dimensionless parameter $\chi$ [cf. Eq.~(\ref{chi})] (for a fixed trapping potential frequency $\Omega$) is sketched for the cases $b_2<0$ and $b_2>0$, respectively. In either case, as $\chi$ changes, the profile of $F_{\rm imp}(\chi)$ remains qualitatively unchanged; for this reason, and for simplicity of the illustration, in Figs.~\ref{veff} and \ref{veff1} we show only one curve for $F_{\rm imp}$. On the other hand, changes of $\chi$ result in a much more pronounced change on $F_{\rm tr}$, since $\chi$ controls its slope. As a result, changes in $\chi$ may lead to the existence of {\it one} or {\it three} fixed points associated with the equation of motion (\ref{eqmot}); the fixed points, $X_0^{\star}$, can be found as solutions of the following transcendental equation:
\begin{equation}
-\omega_{\rm osc}^2 X_0^{\star}+aD_0\sech^2(D_0X_0^{\star})\tanh(D_0X_0^{\star})=0.
\label{fp}
\end{equation}
The possibility of existence of one or three fixed points can also be understood by simplifying the equation of motion (\ref{eqmot}): first, we Taylor expand $V_{\rm eff}(X_0)$ in Eq.~(\ref{pot}) around $X_0=0$ (the location of the impurity) and derive from Eq.~(\ref{eqmot}) a simplified equation of motion for $X_0$ of the following form:
%
\begin{eqnarray}
\ddot{X}_0&=&-\omega_{\rm eff}^2 X_0-K X_0^3,
\label{fp2} \\
\omega_{\rm eff}^2&=&\omega^2_{\rm osc}-\alpha D_0^2,
\label{omeff} \\
K&=&\frac{4}{3}\alpha D_0^4.
\label{K}
\end{eqnarray}
Equation (\ref{fp2}) represents the normal form of the bifurcation
arising in this system; in particular, it indicates that the fixed
point $X_0^*=0$ always exists; nevertheless, depending on $\chi$ and the sign
of $b$, a symmetry-breaking (pitchfork) bifurcation may take place; this way,
two additional fixed points can emerge.
Below we will study the linear stability of the fixed points and obtain characteristic oscillation frequencies $\omega_0$ of small-amplitude motions around them.

\subsection{Attractive impurity}

First, we consider the case of an attractive impurity, $b_2<0$; in this case, one or three fixed points may exist, as shown by the graphical representation of the forces $F_{\rm tr}$ and $F_{\rm imp}$ as functions of the DB-soliton center $X_0$, for different values of the parameter $\chi$ -- see top panel of Fig.~\ref{veff}: it is observed that there exists a critical value of $\chi$, namely $\chi_{c}$, for which $F_{\rm tr}$ is tangent to $F_{\rm imp}$ at $X_0=0$ (see dashed-dotted line in the figure). Then, it can easily be seen that, as long as $\chi <\chi_{c}$ there exists only one fixed point: $X_0^{\star}=0$ (see dotted line in the top panel of Fig.~\ref{veff}). On the other hand, for values $\chi>\chi_{c}$, there exist three fixed points (dashed line of the top panel of Fig.~\ref{veff}). In other words, a typical pitchfork bifurcation occurs at the critical value $\chi_c$: the fixed point at the origin, $X_0^{\star}=0$, loses its stability and, for $\chi>\chi_c$, two new stable (off-center) fixed points emerge. The effective potentials corresponding to the cases $\chi<\chi_c$ and $\chi>\chi_c$ are respectively shown in the left and right bottom panels of Fig.~\ref{veff} and illustrate the symmetry-breaking after the bifurcation.

The above qualitative discussion is also supported by considering the simplified equation of motion (\ref{fp2}), which can also provide some quantitative results for the location and stability of the fixed points, as well as the oscillatory motion of the DB solitons near the fixed points.
Near $X_0=0$, Eq.~(\ref{fp2}) can be approximated by:
\begin{eqnarray}
\ddot{X}_0 \approx -\omega^2_{\rm eff} X_0.
\label{pert1}
\end{eqnarray}
Equation~(\ref{pert1}) describes the motion of a DB-soliton placed near the trap center (where the impurity is located). As long as $\omega_{\rm eff}^2>0$ the soliton will perform small-amplitude oscillations around the center with a frequency $\omega_{\rm eff}$.

Now, if $\chi$ is increased, $\omega_{\rm eff}^2$ (which is positive
for small $\chi$) is decreased and, at the critical point $\chi=\chi_c$, the effective oscillation frequency becomes $\omega_{\rm eff}^2=0$. The critical value $\chi_c$ for which the fixed point $X_0^{\star}=0$ becomes unstable (see dashed-dotted line in the top panel of Fig.~\ref{veff}) can be determined by utilizing Eq.~(\ref{omeff}) -- recall that $\omega_{\rm osc}$, $a$ and $D_0$ in Eq.~(\ref{omeff}) depend on the parameter $\chi$; if   $\chi$ is sufficiently small (an assumption consistent with our previous considerations) then the equation
$\omega_{\rm eff}^2(\chi_c)=0$ leads to the approximate result:
\begin{equation}
\chi_c \approx \frac{2\Omega^2}{\Omega^2-b_2},
\label{xc}
\end{equation}
which is in very good agreement with our numerical findings (see Sec.~IV).
Past this critical point, a soliton placed in the center of the trap will eventually move away from the center and perform large-amplitude oscillations; in this case, $\omega_{\rm eff}^2<0$ and Eq.~(\ref{omeff}) will provide the growth rate of the relevant instability.

As explained above, the symmetry-breaking bifurcation results in the emergence of two new fixed points (see bottom right panel of Fig.~\ref{veff}), which are approximately located at $X_0^{\star} = \pm \omega_{\rm eff}/\sqrt{K}$. The stability of these  nontrivial fixed points can be studied by considering small-amplitude perturbations of Eq.~(\ref{eqmot}), of the form $X_0(t)=X_0^{\ast}+\delta(t)$, and deriving an equation for the small-amplitude perturbations $\delta(t)$:
\begin{eqnarray}
\ddot{\delta}&=&-\omega_0^2 \delta
\label{delta11} \\
\omega_0^2 &=&\omega_{\rm osc}^2 - aD_0^2\sech^2 \left(D_0X_0^{\star} \right)\nonumber \\
&\times & \left[ 3\sech^2 \left( D_0 X_0^{\star}\right) -2 \right].
\label{pert2}
\end{eqnarray}
%
Naturally, this formula applies to an fixed $X_0^{\star}$, including
$X_0^{\star}=0$, in which case it retrieves the result of  Eq.~(\ref{omeff}).

%
%

\subsection{Repulsive impurity}

Let us now consider the case of a repulsive barrier, i.e., $b_2>0$. In this case, the impurity-induced force acting on the DB-soliton is attractive, i.e., $a<0$. Thus, as illustrated in Fig.~\ref{veff1} and also observed from
Eq.~(\ref{fp2}) for $K<0$, the only solution of Eq.~(\ref{fp}) is a trivial fixed point, namely $X_0^{\star}=0$. In this case, we may follow the analysis exposed above and study the stability of $X_0^{\star}=0$, as well as the small-amplitude oscillations around it, by means of Eqs.~(\ref{pert1}) and (\ref{omeff}), but for $a<0$. It is expected that, at least for sufficiently small values of $\chi$, the fixed point should be stable and solitons located near the trap center will perform small-amplitude oscillations. Nevertheless, as will be shown in the next section, the fixed point undergoes an oscillatory instability past a critical value of $\chi$, through a different mechanism.

Below we will compare the above analytical results with numerical simulations.

\section{Numerical results}

In this section, we will numerically investigate the existence of stationary DB-soliton solutions of Eqs.~(\ref{deq1})-(\ref{deq2}), namely $u_d=U_d(x)$ and $u_b=U_b(x)$, located at the fixed points $X_0^{\star}$ obtained before. We will show that such solutions do exist and will subsequently study the linear stability of these states by means of the BdG analysis (see, e.g., Refs.~\cite{emergent,revnonlin,djf}). The latter is performed as follows: we introduce the ansatz
\begin{eqnarray}
u_d(x,t) &=& U_d(x) + \epsilon \left[a(x) e^{i\omega t}
+ b^{\star}(x) e^{-i\omega t} \right],
\label{eq6}
\\
u_b(x,t) &=&  U_b(x) + \epsilon \left[c(x) e^{i\omega t}
+ d^{\star}(x) e^{-i\omega t} \right],
\label{eq7}
\end{eqnarray}
into Eqs.~(\ref{deq1})-(\ref{deq2}), and keeping terms of the order of the small parameter $\epsilon$, we will solve the eigenvalue problem for eigenmodes $\{a(x), b(x), c(x), d(x)\}$ and eigenfrequencies $\omega= \omega_r + i \omega_i$ (note that the stationary state is stable when $\omega_i=0$). This way, we will obtain the excitation spectrum of the relevant stationary states, including characteristic eigenfrequencies associated with the DB-solitons. Such an eigenfrequency is the one pertaining to the ``anomalous mode'' of the system (namely a mode characterized by a negative $energy~\times~norm$ product \cite{book2,fetter}), which coincides with the oscillation frequency of the DB soliton moving near the center of the trap (similarly to the case of dark solitons in one-component BECs \cite{fms}). Following this procedure, we will be able to compare characteristic eigenfrequencies of the excitation spectrum with the oscillation frequencies $\omega_0$ derived in the framework of our analytical approximations. Remarkably, we will show that, generally, there is a very good agreement between the two.

In our numerical results below, we will fix the chemical potential to $\mu=1$, the normalized trap frequency to $\Omega=0.1$, and the impurity strength $b_2=\pm 0.15$ (for the repulsive and attractive cases, respectively). We should also note that, in the numerics, we have approximated the $\delta$-profile of the impurity potential by the function $f(x) = 10\sech^2 (20x)$. Other parameter values produced results qualitatively similar to the ones that will be presented below.

\begin{figure}[tbp]
\centering
\includegraphics[scale=0.4]{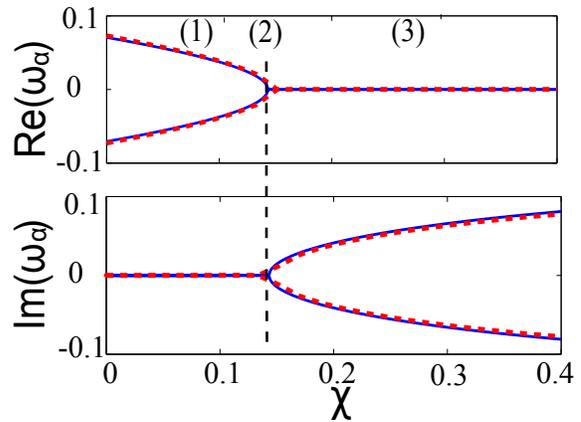}
\caption{(Color online) The top and bottom panels show, respectively, the real part (oscillation frequency) and the imaginary part (instability growth rate) of the anomalous mode eigenfrequency $\omega_a$, as functions of $\chi$, in the case of the (sole) fixed point $X_0^{\star}=0$. Solid (blue) lines indicates $\omega_a$ as obtained from Eq.~(\ref{pert1}), while dashed (red) lines show the numerical result obtained from the BdG analysis. The regimes indicated by (1) and (3) correspond to the cases $\chi<\chi_c$ and $\chi>\chi_c$, while the vertical dashed line (2) indicates the critical value $\chi=\chi_c$.}
\label{bif1}
\end{figure}

\begin{figure}[tbp]
\centering
\includegraphics[scale=0.32]{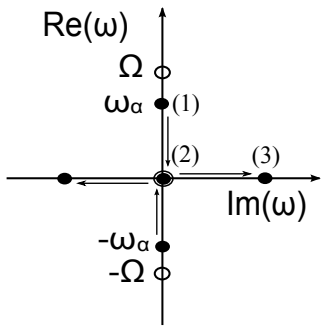}
\caption{A sketch showing the path of the anomalous mode eigenfrequency $\omega_a$ in the excitation spectrum, as the parameter $\chi$ is varied, in the case of the fixed point $X_0^{\star}=0$. As $\chi$ is increased, $\omega_a$ moves towards the zero eigenfrequency of the Goldstone mode, collides with the latter, and an imaginary eigenfrequency pair emerges. The labels (1), (2) and (3) correspond to $\chi<\chi_c$, $\chi=\chi_c$ and $\chi>\chi_c$; see also Fig.~\ref{bif1}.}
\label{sketch1}
\end{figure}
\subsection{Attractive impurity: $b_2<0$ ($a>0$)}

\subsubsection{Fixed point at $X_0^*=0$}

The analytical result of Eq.~(\ref{pert1}), namely the dependence of the oscillation frequency $\omega_0$ on $\chi$, is shown in Fig.~\ref{bif1} [see solid (blue) lines]. On the other hand, in our simulations, we first confirmed the existence of a stationary DB-soliton state located at $x=0$, and then determined its excitation spectrum. The anomalous mode associated with the DB-soliton was found to have an eigenfrequency $\omega_a$, which is almost identical to $\omega_0$ [see dashed (red) lines in Fig.~\ref{bif1}]. Figure~\ref{bif1} clearly illustrates the emergence of the pitchfork bifurcation, occurring
at $\chi_c=0.145$ [see vertical dotted line labeled by (2)]; the regimes (1) and (3) correspond to the cases $\chi<\chi_c$ (one stable fixed point, $X_0^{\star}=0$, in the effective potential) and $\chi>\chi_c$ ($X_0^{\star}=0$ is unstable and two additional fixed points emerge).

In order to better understand the origin of the bifurcation, in Fig.~\ref{sketch1} we show the path of the anomalous mode eigenfrequency $\omega_a$ in the excitation spectrum. At first, i.e., for $\chi=0$, $\omega_a$ is located at $\Omega/\sqrt{2}$, which is the approximate oscillation frequency of dark solitons (in the absence of the bright-soliton component) \cite{djf,fms}. In region (1), $\chi$ is increased and $\omega_a$ moves towards the origin. When $\chi=\chi_c$, $\omega_a$ collides with the zero eigenfrequency of the Goldstone mode -- see region (2) in the figure. This collision gives rise to the emergence of an imaginary eigenfrequency pair, which characterizes the system as long as $\chi>\chi_c$ -- see region (3). The picture shown in Fig.~\ref{sketch1} complements the bifurcation diagrams of Fig.~\ref{bif1}, with the regions (1)-(3) being in correspondence to each other; see also for a discussion of the relevant bifurcation
phenomena in Hamiltonian systems, the recent exposition of~\cite{good}.

\begin{figure}[tbp]
\centering
\includegraphics[scale=0.43]{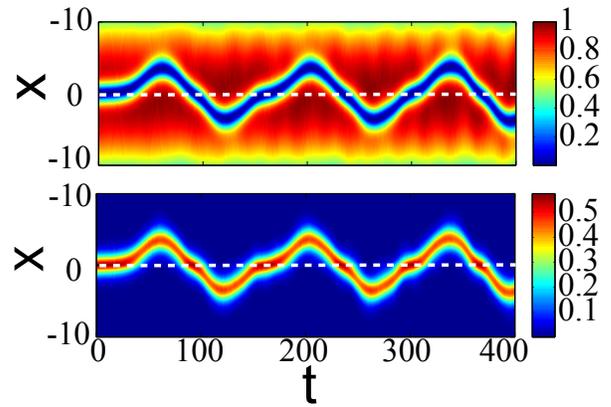}
\caption{(Color online) Contour plot showing the evolution of a DB-soliton, initially placed at $x=0$, for an attractive impurity ($b=-0.15$), and for $\chi=1.35$ (in this case, the fixed point $X_0^{\star}=0$ is unstable). Top and bottom panels show the dark- and bright-soliton components, respectively. The dashed (white) line indicates the location of the impurity.}
\label{dyn1}
\end{figure}

We have also studied numerically the manifestation of the above mentioned instability of a stationary DB-soliton (initially located at $x=0$), by using this state as initial condition, and numerically integrating Eqs.~(\ref{deq1})-(\ref{deq2}).
Note that to trigger the onset of the instability, a small random  perturbation [of $O(10^{-3})$)] was added to the initial condition. The result is illustrated in Fig.~\ref{dyn1}, where the time evolution of a stationary DB-soliton is shown, for $\chi=1.35>\chi_c$. As seen in the figure, the initially stationary DB-soliton is exponentially unstable
and eventually departs from its initial location, and starts performing oscillations. Notice that the soliton energy is sufficiently large so that the soliton is always transmitted through the effective barrier located at the origin. It is clearly observed that the interaction of the soliton with the impurity results in a position shift: in fact, as the soliton moves from the one well of the effective double-well potential (see bottom right panel of Fig.~\ref{veff}) to the other, it slows down at the impurity for a short time and, afterwards, it is transmitted to the other well.

\subsubsection{Fixed points at the minima of the effective double-well potential}

\begin{figure}[tbp]
\centering
\includegraphics[scale=0.4]{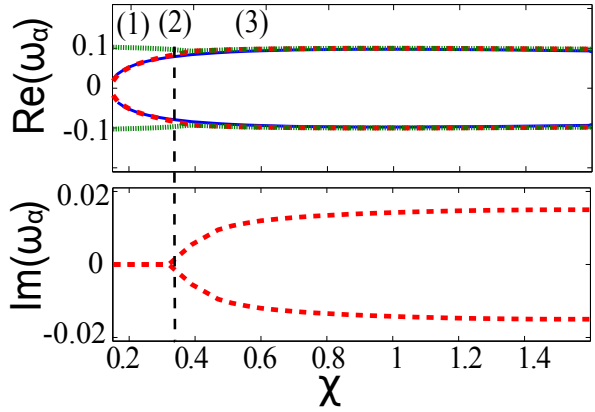}
\caption{(Color online) Same as Fig.~\ref{bif1}, but in the case of the nontrivial fixed points $X_0^{\star}$. Solid (blue) lines indicates $\omega_a$ as obtained from Eq.~(\ref{pert2}), dashed (red) lines show the numerical result obtained from the BdG analysis, while dotted (green) line in the top panel of the figure indicates the eigenfrequency of the approximate Kohn mode. The regimes indicated by (1), (2) and (3) correspond to the cases $\chi<\chi_{c1}$, $\chi=\chi_{c1}$ and $\chi>\chi_{c1}$.}
\label{bif2}
\end{figure}

\begin{figure}[tbp]
\centering
\includegraphics[scale=0.32]{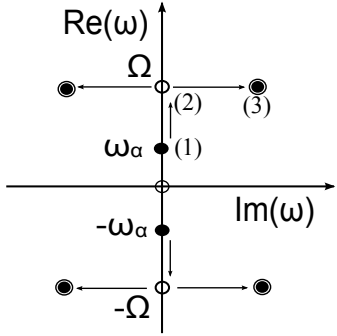}
\caption{Similar to Fig.~\ref{sketch1}, but in the case of the nontrivial fixed points. As the parameter $\chi$ is increased, the anomalous mode eigenfrequency $\omega_a$ moves towards the Kohn mode eigenfrequency (located at $\omega=\Omega$) and, after the collision, a complex eigenfrequency quartet emerges. The regimes indicated by (1), (2) and (3) correspond to the cases $\chi<\chi_{c1}$, $\chi=\chi_{c1}$ and $\chi>\chi_{c1}$; see also Fig.~\ref{bif2}.}
\label{sketch2}
\end{figure}

As in the case of $X_0^{\star}=0$, we numerically confirmed the existence of stationary DB-soliton states located at the nontrivial fixed points, and then determined their excitation spectra. In Fig.~\ref{bif2}, we compare the result of Eq.~(\ref{pert2}) with the one obtained in the framework of BdG analysis. An excellent agreement between the two is observed, up to a critical value of $\chi$, namely $\chi_{c1}=0.32$: in this regime, $\omega_0$ of Eq.~(\ref{pert2}) [solid (blue) line in the top panel of Fig.~\ref{bif2}] coincides with the real part of the anomalous mode eigenfrequency $\omega_a$ [dashed (red) line]. Nevertheless, at $\chi=\chi_{c1}$, the BdG analysis reveals that $\omega_a$ collides with the eigenfrequency  $\omega\approx\Omega$, [the so-called Kohn (or dipolar) mode
for $b=0$], which characterizes the TF background \cite{book2}. This collision results in the emergence of an unstable excitation mode, characterized by a
complex eigenfrequency quartet, the imaginary part of which are shown in the bottom panel of Fig.~\ref{bif2}. In this case, a Hamiltonian-Hopf bifurcation
takes place. This procedure can be better understood in the sketch shown in Fig.~\ref{sketch2}: as the parameter $\chi$ is increased, the anomalous mode eigenfrequency $\omega_a$ is also increased, i.e., it moves to the opposite direction as compared to the situation shown in Fig.~\ref{sketch1}. This way, $\omega_a$ eventually collides with the eigenfrequency $\omega\approx\Omega$ , and gives rise to the emergence of a quartet of complex eigenfrequencies.

\begin{figure}[b]
\centering
\includegraphics[scale=0.4]{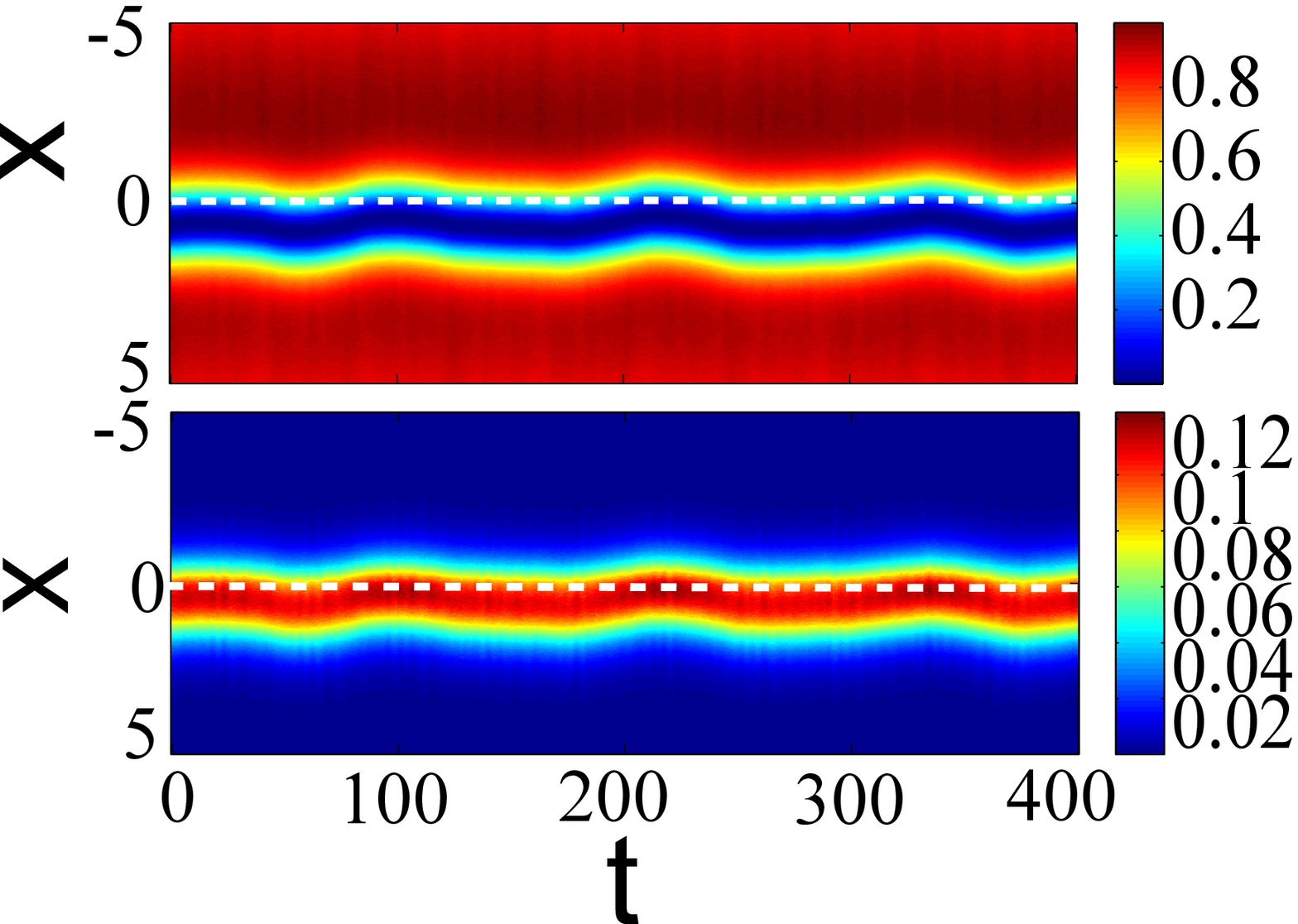}
\includegraphics[scale=0.4]{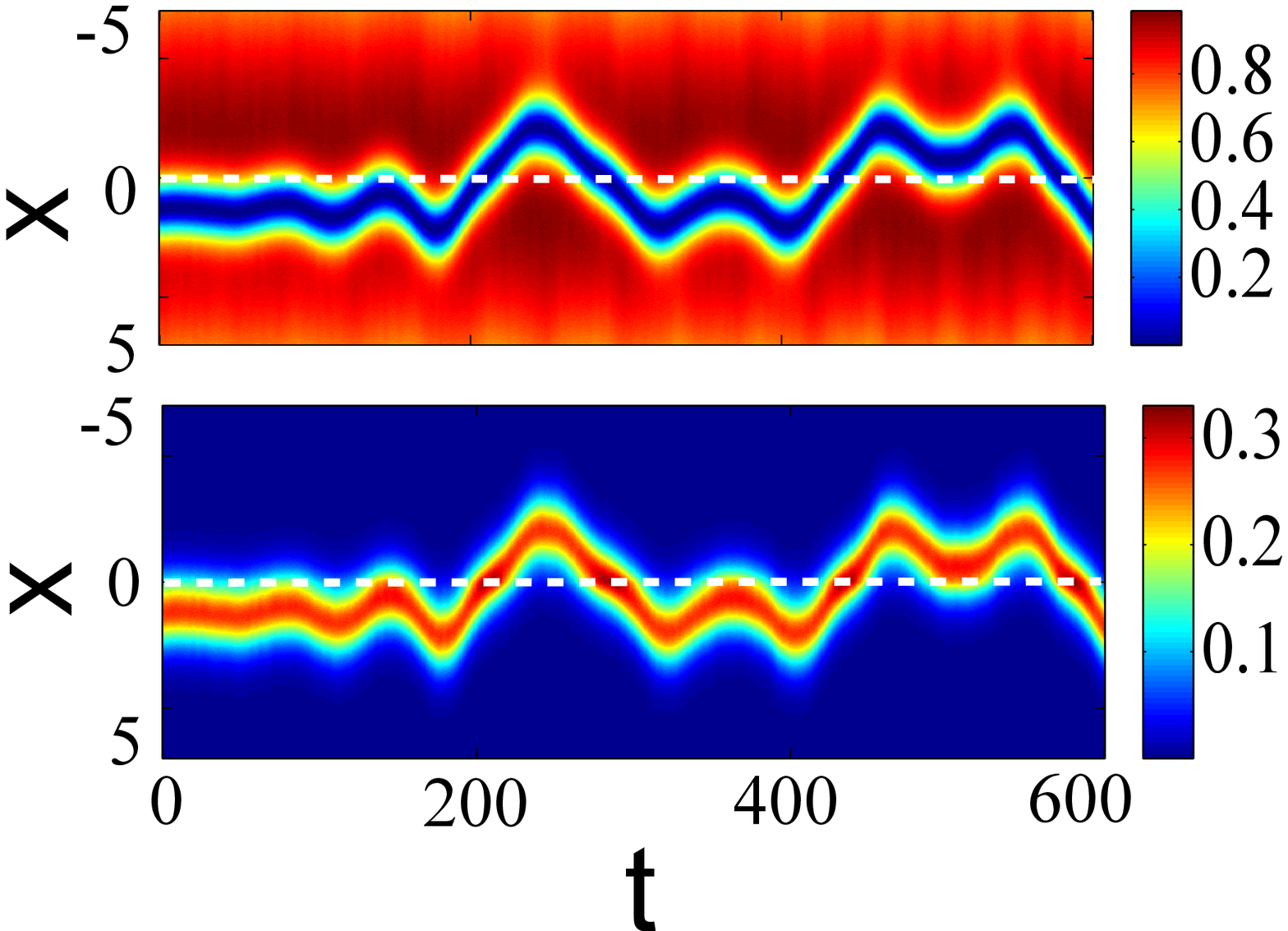}
\caption{(Color online) The two top panels present contour plots showing the evolution of a DB-soliton, initially placed at the fixed point $x=X_0^{\star}=0.6$, for a value of $\chi=0.23<\chi_{c1}=0.32$; in this case, the fixed point is stable. The two bottom panels are similar to the two top ones, but for the fixed point $x=X_0^{\star}=1.3$, for a value of $\chi=0.62>\chi_{c1}$; in this case, the fixed point is oscillatorily unstable. First- and third- (second- and fourth-) row panels show the dark- (bright-) soliton components. The dashed (white) line indicates the location of the impurity.}
\label{dyn2}
\end{figure}

The above analysis suggests that for values $\chi<\chi_{c1}$, a DB-soliton initially located at any of the two nontrivial fixed points, when displaced,  will perform small-amplitude oscillations at one well of the effective double-well potential. A direct numerical integration of Eqs.~(\ref{deq1})-(\ref{deq2}), with initial condition such a stationary DB-soliton state (perturbed by random noise), shows that this is the case indeed: a prototypical example is shown in two top panels of Fig.~\ref{dyn2}, where the dynamics of such a state is illustrated, for $\chi=0.25<\chi_{c1}=0.32$. It is clearly observed that the DB-soliton oscillates around the center of one of the wells, with an oscillation frequency $\omega_a \approx 0.05$; this value deviates approximately $5\%$ from the analytically predicted value [cf. Eq.~(\ref{pert2})]. On the other hand, it is interesting to numerically investigate the manifestation of the predicted instability of a stationary DB-soliton state for $\chi>\chi_{c1}$. Such a case, is illustrated in the two bottom panels of Fig.~\ref{dyn2}, where the evolution of such a DB-soliton is shown, for $\chi=0.62$. It is observed that the initially quiescent DB-soliton starts performing small-amplitude oscillations around the center of one of the wells but, after a short time, it gains enough kinetic energy to be transmitted through the effective barrier. This way, it moves over to the other well of
the effective double-well potential and, afterwards, the above process is repeated.


\subsection{Repulsive impurity: $b_2>0$ ($a<0$) }

\begin{figure}[tbp]
\centering
\includegraphics[scale=0.4]{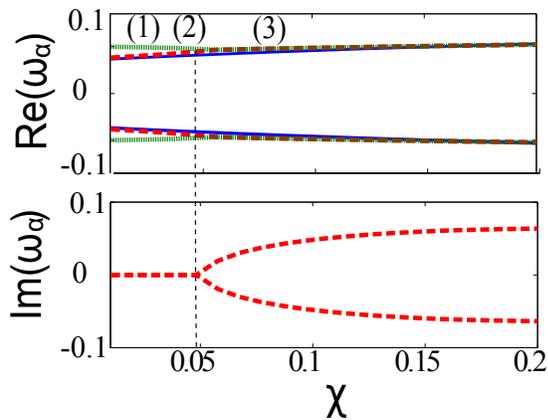}
\caption{(Color online) Similar to Fig.~\ref{bif2}, but for the case of a repulsive impurity ($b_2=0.15$) and for a DB-soliton located at $x=0$. The critical value of parameter $\chi$ is $\chi_{c2}=0.05$.}
\label{bif3}
\end{figure}

In the case of a repulsive barrier impurity, we will compare the relevant analytical [see Eq.~(\ref{pert1}) for $a<0$] and numerical results (obtained by the BdG analysis). First we mention that, as seen in the top panel of Fig.~\ref{bif3}, the oscillation frequency $\omega_0$ of the DB-soliton almost coincides with the anomalous mode eigenfrequency $\omega_a$, only for sufficiently small values of parameter $\chi$. In fact, there exists a critical value of $\chi$, namely $\chi_{c2} = 0.05$, where a bifurcation -- similar to the one shown in Fig.~\ref{sketch2} -- takes place. This bifurcation results in the emergence of an unstable eigenmode, characterized by a quartet of complex eigenfrequencies, the imaginary part of which is shown in the bottom panel of Fig.~\ref{bif3}.

\begin{figure}[b]
\centering
\includegraphics[scale=0.4]{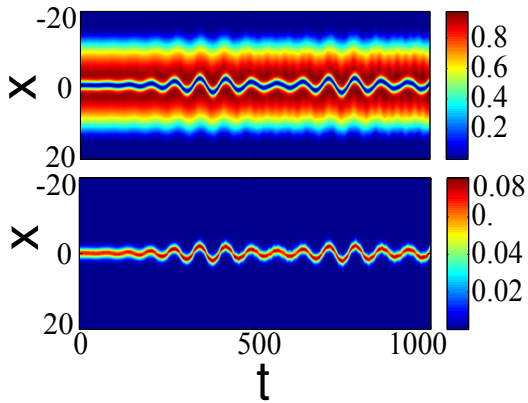}
\caption{(Color online) Similar to Fig.~\ref{dyn1}, but for the case of a repulsive impurity ($b_2=0.15$).}
\label{dyn3}
\end{figure}

As before, it is relevant to numerically study the manifestation of the instability in the case of a DB-soliton initially placed at $x=0$, for $\chi>\chi_{c2}$. A pertinent example is illustrated in Fig.~\ref{dyn3}, where the evolution of such a state is shown for $\chi=0.15$. The soliton falls into an instability, and eventually starts to oscillate around the center of the trap.


\section{Conclusions}

We have used mean-field theory to study the statics and dynamics of atomic dark-bright solitons in the presence of localized (delta-like) impurities. Our model considered a system of two coupled Gross-Pitaevskii equations, describing a two-component Bose-Einstein condensate, confined in an external potential composed of a harmonic trap and a pair of localized impurities acting on each component.

We have employed the adiabatic perturbation theory for solitons to derive an equation of motion for the dark-bright soliton center. Our analytical approximation revealed that if both impurity potentials are repulsive (attractive) then the effective potential felt by the soliton is either a double-well or a harmonic potential with a localized well located in the trap center. Investigating the forces acting on the soliton, we have identified an interesting situation, which was then analyzed in detail: if the impurity potential acts solely on the bright-soliton component, then the impurity-induced part of the effective potential is either a localized barrier (for attractive impurity) or a localized well (for repulsive impurity). This behavior is in sharp contrast with the one corresponding to the case of single-component condensates, where the nature (repulsive or attractive) of the soliton-impurity interaction is identical to the type of the impurity (repulsive or attractive) \cite{gt}.

Our numerical simulations have confirmed that stationary dark-bright solitons do exist at the fixed points of the effective potential. The stability of these fixed points was studied, and the frequency of
small-amplitude oscillations in a stable configuration was found analytically. We have performed a Bogoliubov-de Gennes analysis to study the stability of stationary states and find their excitation spectrum. The eigenfrequencies of the anomalous modes were found to be almost identical to the analytically obtained soliton oscillation frequencies, at least for sufficiently small number of atoms of the bright component. In the case of unstable fixed points, the bifurcations
(pitchfork or Hamiltonian Hopf) that give rise to the destabilization
are identified and the growth rate of the perturbations are theoretically
identified and corroborated by numerical linear stability analysis.

An interesting direction for future studies would be a systematic study of the scattering of dark-bright solitons from localized impurities of arbitrary amplitude (in the lines of the work in Refs.~\cite{kos2,cao,holmes,vvk1,leebrand,brand2}). Furthermore, it would be interesting to study similar problems but for impurities that have spatial scales larger than the ones of the soliton, and investigate possible changes in the stability and dynamics. Additionally, it would be quite relevant to extend the present analysis (and its pertinent generalizations as per the previous points) to multi-dimensional settings, and study the statics and dynamics of vortices in the presence of localized impurities (see, e.g., a relevant study but for a single-component condensate in Ref.~\cite{ricpanos}). Such studies are in progress and pertinent results will be presented elsewhere.

\section*{Acknowledgments} This work, initiated by constructive discussions with Evgeny V. Doktorov who recently passed away, is dedicated to his memory. The work of D.J.F. was partially supported by the Special Account for Research Grants of the University of Athens. The work of P.G.K. was supported by the National Science Foundation Grant NSF-DMS-0806762, as well as by the Alexander von Humboldt Foundation. The authors are also grateful to Peter Engels, for numerous discussions on dark-bright solitons and helpful comments on the manuscript.


\appendix
\section{Equation of motion for the soliton center}

Substituting $R_d$ and $R_b$ [cf.~Eqs.~(\ref{Rd})-(\ref{Rb})] into Eq.~(\ref{perturb}) and evaluating the integrals, we obtain from
Eq.~(\ref{perturb}) the following result:
\begin{eqnarray}
\frac{dE}{dt}&=&\mu^{-2} \sin(2\phi) \left(\cos^2\phi-\eta^2 \right) V'(x) \nonumber \\
&-&\frac{1}{2}b_1 D \sin(2\phi) (\cos^2\phi+2D^2)I_1
\nonumber \\
&+& b_2 \chi \mu^{-2} D^3 \tan\phi \sech^2(Dx_0)\tanh(Dx_0) \nonumber \\
&+& \frac{1}{2} b_1 \chi \mu^{-2} D^2 \tan\phi (I_1 \cos^2\phi-I_2),
\label{fulleq2}
\end{eqnarray}
where we have Taylor expanded the potential $V(x)$ around the soliton center $x_0$ and assumed that the DB-soliton is moving in the vicinity of the trap center (where the impurity is located), $x_0 \approx 0$; this way, we actually deal with nearly stationary DB-solitons, characterized by slow velocities, such that the phase angle is $\phi \approx 0$. Furthermore, $I_1$ and $I_2$ in Eq.~(\ref{fulleq2}) are the following integrals:
\begin{eqnarray}
I_1 &=& \int_{-\infty}^{+\infty} \left[\frac{x}{|x|}\sech^4[D(x-x_0)]{\rm e}^{-2|x|}\right] dx,
\label{I1} \\
I_2 &=& \int_{-\infty}^{+\infty} \left[\frac{x}{|x|}\sech^2[D(x-x_0)]{\rm e}^{-2|x|}\right] dx,
\label{I2}
\end{eqnarray}
which can be evaluated by means of the hypergeometric functions \cite{abram}. Nevertheless, in the physically relevant case of sufficiently small $\chi$ [cf. Eq.~(\ref{chi})], i.e., when the number of atoms of the bright soliton is only a small fraction of the total number of atoms \cite{hamburg,engels1,engels2,engels3}, we may approximate the above integrals as $I_1 \approx I_2 \approx (2/3)\sech^2(Dx_0)\tanh(Dx_0)$. This way, we accordingly simplify Eq.~(\ref{fulleq2}), which together with Eqs.~(\ref{s1}), (\ref{s2}) [and Eq.~(\ref{denergy1})] constitute a system of three ordinary differential equations for the unknown soliton parameters $\phi(t)$, $x_0(t)$ and $D(t)$. This system can be solved approximately
upon linearizing around the fixed point:
\begin{equation}
\phi_0=0, \,\,\, x^{(0)}_0 =0, \,\,\,
D_0=\sqrt{1+\left(\frac{\chi}{4}\right)^2}-\frac{\chi}{4},
\label{fixedpoint}
\end{equation}
using the ansatz $x_0=X_0$, $\phi = \phi_1$ and $D=D_0+D_1$. We thus obtain the following results:
%
\begin{eqnarray}
D_1&=&-\tilde{D}_0\phi_1^2, \quad \tilde{D}_0 \equiv \left(2D_0+\frac{\chi}{2}\right)^{-1},
\label{D1} \\
\mathcal{D} \dot{\phi}_1&=&-2 +\chi D_0 - D_0 \nonumber \\
&\times& \left[ \frac{2b_1}{3}(1+2D_0^2) - \chi D_0 \left(b_2 D_0 - \frac{b_1}{3}\right) \right] \nonumber \\
&\times& {\rm sech}^2(D_0 X_0)\tanh(D_0 X_0),
\label{phidot} \\
\dot{X}_0&=&D_0 \phi_1,
\label{dotx}
\end{eqnarray}
where
\begin{eqnarray}
\mathcal{D}= -D_0 \left[ 8^2\tilde{D}_0+ \chi(2\tilde{D}_0-D_0)\right].
\label{cald}
\end{eqnarray}
To this end, differentiating Eq.~(\ref{dotx}) with respect to time once, and using Eq.~(\ref{phidot}), after some straightforward algebraic manipulations, we obtain the equation of motion (\ref{eqmot}) for the DB soliton center.

\end{document}